\documentclass[english,superscriptaddress,reprint,two column]{revtex4-1}
\usepackage[T1]{fontenc}
\usepackage{color}
\usepackage{amsmath}
\usepackage{amssymb}
\usepackage{graphicx}
\usepackage{amsfonts}
\usepackage{transparent}
\usepackage{babel}
\usepackage{bm}
\usepackage{epsfig}
\usepackage{enumerate}
\usepackage{multirow}
\usepackage{verbatim}
\usepackage{float}
\usepackage[breaklinks=true,colorlinks=true,linkcolor=blue,urlcolor=blue,citecolor=blue]{hyperref}
\usepackage{caption}


\renewcommand{\raggedright}{\leftskip=0pt \rightskip=0pt plus 0cm}
\begin{document}

\title{Perfect nonreciprocity by loss engineering}

\author{Xinyao Huang}
\affiliation{State Key Laboratory of Low-Dimensional Quantum Physics, Department of
Physics, Tsinghua University, Beijing 100084, China}
\author{Yong-Chun Liu}
\email{ycliu@tsinghua.edu.cn}
\affiliation{State Key Laboratory of Low-Dimensional Quantum Physics, Department of
Physics, Tsinghua University, Beijing 100084, China}
\affiliation{Frontier Science Center for Quantum Information, Beijing 100084, China}

\begin{abstract}
Realization of nonreciprocal transmission with low insertion loss and high contrast simultaneously is in great demand for one-way optical communication and information processing. 
Here we propose a generic approach to achieving perfect nonreciprocity that allows lossless unidirectional transmission by engineering energy losses. 
The loss of the intermediate mode induces a phase lag impinging on the indirect channel for energy transmission, which does not depend on the energy transmission direction. 
When the direct transmission channel coexists with the indirect lossy transmission channel, the dual-channel interference can be tuned to be  destructive simultaneously for backward transmission from the rightmost mode to the leftmost mode, and for forward transmission from the leftmost mode to the intermediate mode. The former interference outcome corresponds to 100\% nonreciprocity contrast, and the latter guarantees zero insertion loss for forward transmission. 
Additionally, our scheme also allows a nonreciprocity response over a wide bandwidth by increasing the losses while keeping perfect nonreciprocity at resonance. The robustness against loss indicates that our scheme is advantageous in the implementation of nonreciprocal optical devices with high performance.
\end{abstract}

\maketitle

\section{INTRODUCTION} 

As a key ingredient in optical information processing, optical nonreciprocity that allows optical fields to only propagate in one way  offers an efficient means to route the signal unidirectionally and protect the source from backscattering dissipation~\cite{Jalas2013,Caloz2018,Asadchy2020}. 
The most conventional approach to nonreciprocity generation is based on magneto-optical effects to break the Lorentz reciprocity theorem~\cite{Haldane2008, Hadad2010, Khanikaev2010,Bi2011}. However, due to the use of an external magnetic field and material losses, it is challenging to use magnetic-based schemes to integrate  nonreciprocal devices on chip with low insertion loss~\cite{Dai2012}. To overcome this challenge, various magnetic-free protocols have been proposed, such as methods based on spatiotemporal modulation of system permittivity~\cite{Yu2009, Kang2011, Tzuang2014, Estep2014,Koutserimpas2018,Kittlaus2018},  optical nonlinearity~\cite{Khanikaev2015,Shi2015}, optomechanical interactions~\cite{Kim2015,Shen2016, Ruesink2016,Peterson2017,Barzanjeh2017,Metelmann2015,Fang2017,Xu2019}, the Sagnac effect induced by spinning resonators~\cite{Maayani2018, Huang2018},  and  atomic thermal motion~\cite{Xia2018, Zhang2018, Liang2020}. However nonreciprocal transmission with low insertion loss and high contrast simultaneously has remained elusive due to reasons such as the fundamental limitation of the maximum forward transmission for nonlinear single-resonator devices~\cite{Sounas2018}, stringent requirements of system parameters for spinning resonators ~\cite{Maayani2018}, and so on. Although recent schemes based on the use of cascaded nonlinear resonators~\cite{Yang2020} or feedback control~\cite{Tang2021} have been applied to improve the performance of the nonreciprocal transmission, achieving perfect nonreciprocity that allows lossless field transmission one way (i.e., zero insertion loss) and blocks field transmission in the opposite direction (i.e., 100\% contrast) simultaneously is still an outstanding challenge.
 
Here we propose an efficient scheme for generating perfect nonreciprocity by engineering energy losses.
Our scheme works for generic bosonic oscillators that can be implemented in a wider range of systems, 
 such as optical cavities~\cite{Vahala2003}, exciton polaritons~\cite{Mandal2020}, micromechanical oscillators~\cite{Aspelmeyer2014,Riedinger2018, Ockeloen-Korppi2018}, and ensembles of atoms~\cite{Hammerer2010}. 
The essential mechanism of our scheme is that completely opposite interference outcomes for forward and backward energy transmission can be achieved simultaneously. This hinges on the dual-channel transmission, wherein the energy can be transferred either from the direct coherent coupling channel or from the indirect lossy coupling channel implemented by an intermediate mode with nonzero energy loss. 
Based on the scheme proposed in Ref.~\cite{Huang2021}, the phase lag induced by the energy loss is independent of the transmission direction. When more than one lossy coupling channel exists, the interference outcomes between different lossy coupling channels can be tuned to be different for the forward and backward directions, leading to nonreciprocal transmission. Due to the existence of the energy losses in the two lossy coupling channels, the unidirectional transmission efficiency can only be maximized to 68.6\% by optimizing the system parameters. However, when one lossy coupling channel is replaced by a direct coherent coupling channel,
simultaneous destructive interference for backward (forward) energy transmission  from the rightmost (leftmost)  mode to the leftmost (rightmost) mode, and for forward (backward) energy transmission from the leftmost (rightmost) mode to the intermediate mode can be obtained. The former interference outcome corresponds to the unidirectional forward (backward) energy transmission with 100\% nonreciprocity contrast, and the latter ensures the complete forward (backward) energy transmission to the rightmost (leftmost) mode, indicating zero forward (backward) insertion loss. Additionally, we find that the nonreciprocity bandwidth can be broadened by increasing the losses of the resonant modes while keeping the perfect nonreciprocity at resonance.

This paper is organized as follows: In Sec.~II, we describe the theoretical model for a system composed of coupled resonance modes with energy losses and then, in Sec.~III, we demonstrate how to generate perfect nonreciprocity by engineering losses and dual-channel interference. In Sec.~IV, we show that the perfect nonreciprocity realized in our scheme is robust against energy loss and calculate the nonreciprocity bandwidth. Finally, we summarize our results and discuss possible experimental implementations in Sec.~V.

\section{Model} \label{sec: model}

\begin{figure}
\includegraphics[width=\columnwidth]{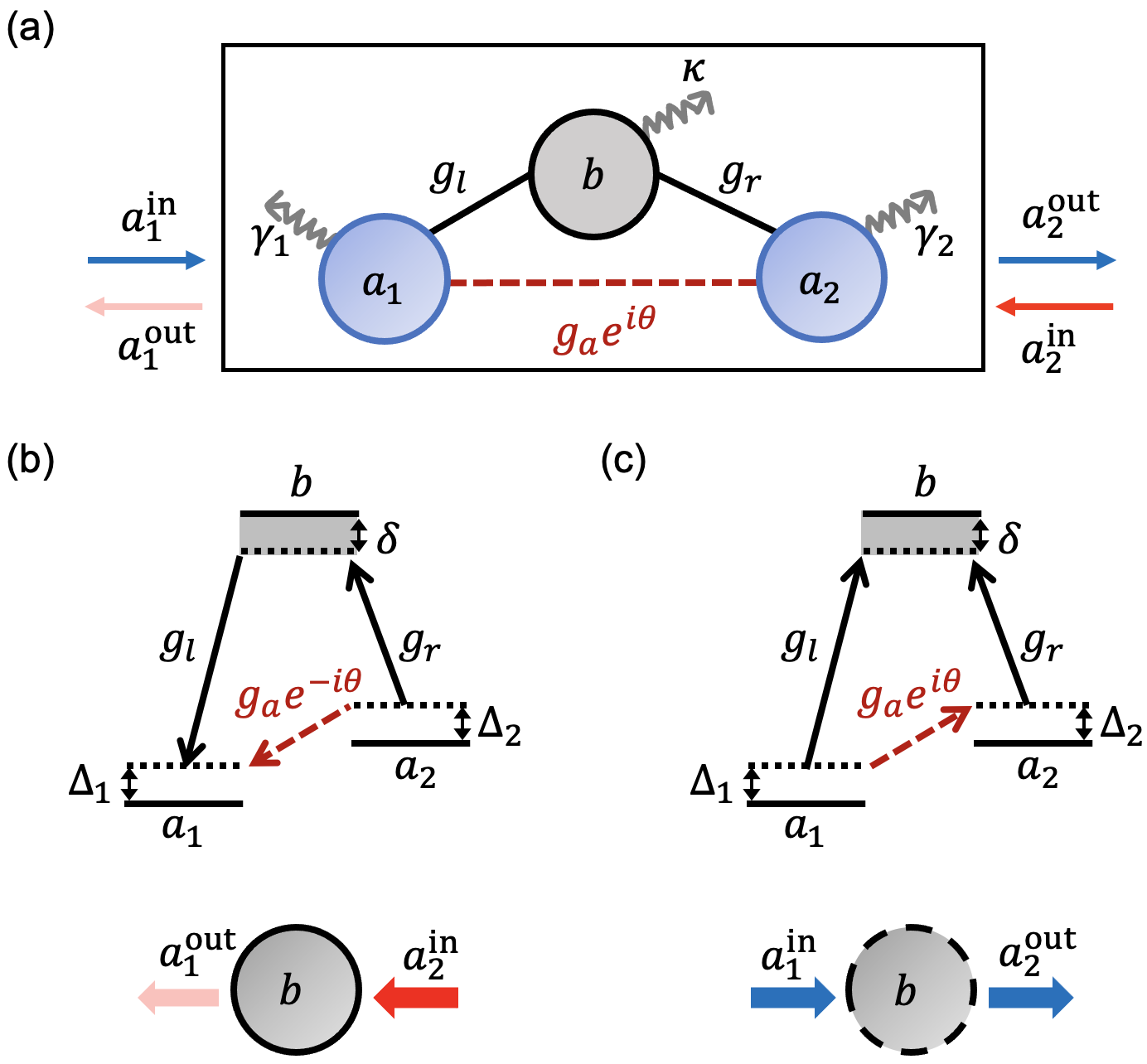}
\caption{(a) A system composed of two resonance modes $a_{1}$ and $a_{2}$  coupled with a lossy mode $b$.  The coupling constant is denoted by  $g_{l}$ ($g_{r}$). $a_{1}$ and $a_{2}$ are also coupled with each other directly. The coupling coefficient is represented as the constant $g_{a}$ with a phase factor $e^{i\theta}$. The  energy decay rates of the resonance modes $a_{1}$, $a_{2}$, and $b$ are represented as $\gamma_{1}$, $\gamma_{2}$, and $\kappa$, respectively. 
 (b) and (c) Energy-domain illustration of dual-channel interference for implementing perfect nonreciprocity. The solid horizontal lines represent the resonance frequencies of the modes, and the dashed horizontal lines are respective detunings.  
 Engineering the interference of the two channels for backward energy transmission (b), i.e.,  for direct channel $a_{2}\rightarrow a_{1}$  and indirect channel $a_{2}\rightarrow b\rightarrow a_{1}$, to be destructive, zero energy occupation on $a_{1}$ warrants that input from $a_{2}$ cannot be transmitted to $a_{1}$, indicating the blockade of backward energy transmission. Applying this interference mechanism to the forward energy transmission (c), when the interference between the direct channel $a_{1}\rightarrow b$  and indirect channel $a_{1}\rightarrow a_{2}\rightarrow b$  is tuned to be destructive, zero energy occupation on $b$ makes it transparent for forward transmission ($a_{1}\rightarrow a_{2}$), indicating zero insertion loss for forward transmission. Hence lossless unidirectional forward energy transmission can be implemented by combining the interference mechanism depicted in (b) and (c).}
\label{fig1}
\end{figure}

As illustrated in Fig.~\ref{fig1}, we consider a generic system composed of  two resonance modes $a_{1}$ and $a_{2}$ with direct interaction coupled to a lossy mode $b$. The system Hamiltonian  is given as ($\hbar=1$)

\begin{equation}
\begin{aligned}
H_{\text{eff}}=&-\Delta_{1}a_{1}^{\dagger}a_{1}-\Delta_{2}a_{2}^{\dagger}a_{2}-\delta b^{\dagger}b\\
&+(g_{l}a_{1}b^{\dagger}+g_{r}a_{2}b^{\dagger}+g_{a}e^{-i\theta}a_{1}^{\dagger}a_{2})+\text{H.c.},
\label{eq:effH}
\end{aligned}
\end{equation}
where $\Delta_{1}$, $\Delta_{2}$, and $\delta$ are the detunings with respect to the resonance frequencies of each mode under the rotating frame.  The direct coupling between $a_{1}$ and $a_{2}$ contains a nonzero coupling phase $\theta$, which is typically implemented by using nonlinearity and time modulation (see details in Appendix A).
Defining a vector $\vec{v}=(a_{1}, b, a_{2})^{\text{T}}$ in terms of the annihilation operators of the modes, the Langevin equations of $\vec{v}$ can be written as $(d\vec{v}/dt)=M\vec{v}+\sqrt{\Gamma_\mathrm{ex}}\vec{v}_\mathrm{in}+\sqrt{\Gamma_{0}}\vec{f}_\mathrm{in}$,
 where the coefficient matrix is
\begin{equation}
 M=\begin{pmatrix} i\Delta _{1}-\gamma _{1}/2&&-ig_{l}&&-ig_{a}e^{-i\theta}\\-ig_{l}&&i\delta-\kappa/2&&-ig_{r}\\-ig_{a}e^{i\theta}&&-ig_{r}&&i\Delta _{2}-\gamma _{2}/2 \end{pmatrix}.
\end{equation}
$\vec{v}_\mathrm{in}=(a_{1}^\mathrm{in}, b_\mathrm{in}, a_{2}^\mathrm{in})^{\text{T}}$ is the input field vector, with its associate decay rate $\sqrt{\Gamma_\mathrm{ex}}=\text{Diag}[\sqrt{\gamma_{1}^\mathrm{ex}},\sqrt{\kappa_\mathrm{ex}}, \sqrt{\gamma_{2}^\mathrm{ex}}]$. $\vec{f}_\mathrm{in}$ accounts for the  additional vacuum noise fields due to the intrinsic dissipation rate $\sqrt{\Gamma_{0}}=\text{Diag}[\sqrt{\gamma_{1}^{(0)}},\sqrt{\kappa_{0}},\sqrt{\gamma_{2}^{(0)}}]$.
Using the Fourier transform convention $\vec{v}(t)=\int d\omega e^{-i\omega t}\vec{v}(\omega)/(\sqrt{2\pi})$, the steady-state solution of the field amplitude can be expressed in terms of the input field operators as
\begin{equation}
\begin{aligned}
a_{1}(\omega)\approx \sqrt{\gamma_{1}}A_{1}a_{1}^{\text{in}}+\sqrt{\gamma_{2}}A_{2} a_{2}^{\text{in}},\\
b_{1}(\omega)\approx \sqrt{\gamma_{1}}B_{1}a_{1}^{\text{in}}+\sqrt{\gamma_{2}}B_{2} a_{2}^{\text{in}},\\
a_{2}(\omega) \approx \sqrt{\gamma_{1}}C_{1}a_{1}^{\text{in}}+\sqrt{\gamma_{2}}C_{2} a_{2}^{\text{in}},
\end{aligned}
\label{eq:ssol}
\end{equation}
where we ignore the intrinsic dissipation ($\gamma_{i}=\gamma_{i}^{\mathrm{ex}}+\gamma_{i}^{(0)}\approx\gamma_{i}^{\mathrm{ex}}$, $\kappa=\kappa_{\mathrm{ex}}+\kappa_{0}\approx\kappa_{\mathrm{ex}}$) and assume the input fields coupled with the modes $a_{1}$ and $a_{2}$, i.e., $b_{\text{in}}=0$. The coefficients $A_{i}$, $B_{i}$, and $C_{i}$, $i\in\{1,2\}$, are functions of the coupling constants ($g_{l}, g_{r}, g_{a}$), phase ($\theta$), decay rates ($\gamma_{1}$, $\gamma_{2}$, $\kappa$), and detunings ($\Delta_{1}$, $\Delta_{2}$, $\delta$) (the detailed expression can be found in Appendix B).

Plugging Eq.~\eqref{eq:ssol} into the input-output relation $\vec{v}_\mathrm{out}=\vec{v}_\mathrm{in}-\sqrt{\Gamma_{\mathrm{ex}}}\vec{v}$, we can derive the forward ($a_{1}\rightarrow a_{2}$) and backward ($a_{1}\leftarrow a_{2}$) energy transmission efficiencies as (see derivation in Appendix B)
\begin{equation}
\begin{aligned}
T_{\rightarrow}(\omega)=|\langle a_{2}^{\text{out}}/a_{1}^{\text{in}}\rangle|^{2}=\gamma_{1}\gamma_{2}|C_{1}|^{2},\\
T_{\leftarrow}(\omega)=|\langle a_{1}^{\text{out}}/a_{2}^{\text{in}}\rangle|^{2}=\gamma_{1}\gamma_{2}|A_{2}|^{2},
\label{eq:Teff}
\end{aligned}
\end{equation}
where
\begin{equation}
\begin{aligned}
C_{1}&\propto g_{l}g_{r}+g_{a}|\Omega_{b}|e^{i(\theta+\phi_{b})},\\
A_{2}&\propto g_{l}g_{r}+g_{a}|\Omega_{b}|e^{-i(\theta-\phi_{b})}.
\label{eq:coeff}
\end{aligned}
\end{equation}
Here, $\Omega_{b}=\delta+\omega+i\kappa/2$ is the effective resonance frequency of the lossy mode $b$.
Equation~\eqref{eq:Teff} and~\eqref{eq:coeff} indicate that the energy input from the port $a_{1}^{\mathrm{in}}$ ($a_{2}^{\mathrm{in}}$) can be transferred to $a_{2}$ ($a_{1}$) along two possible channels due to the coexistence of the direct coupling and indirect coupling mediated by the lossy mode $b$. Besides the nonzero coupling phase $\theta$, a phase lag $\phi_{b}=\arg(\Omega_{b})$ induced by the nonzero energy loss ($\kappa\neq0$) of $b$ will be added on the indirect channel for energy transmission ($a_{1}\leftrightarrow b\leftrightarrow a_{2}$). Since  $\phi_{b}$ is independent of the transmission direction, the direct and indirect paths will interfere differently for forward and backward transmission, leading to unequal transmission efficiencies $T_{\rightarrow}(\omega)\neq T_{\leftarrow}(\omega)$, i.e., nonreciprocal energy transmission. 

Zero transmission efficiency corresponds to tuning the coupling strengths and phases to satisfy the conditions
\begin{equation}
\begin{aligned}
g_{l}g_{r}&= g_{a}|\Omega_{b}|,\\
\theta\mp\phi_{b}&=(2k+1)\pi,\\
\theta\neq p\pi, &\quad \phi_{b}\neq q\pi,
\end{aligned}
\label{eq:uni-cond}
\end{equation}
where $k,p,q$ are integers. In this case, unidirectional forward ($-$) or backward ($+$) transmission can be achieved by engineering the destructive interference of the two channels to block the energy transmission from the opposite direction.
 As shown in Fig.1(b), unidirectional forward transmission is implemented as the destructive interference of the direct channel ($a_{1}\leftarrow a_{2}$) and indirect channel ($a_{1}\leftarrow b\leftarrow a_{2}$) blocks the energy transfer from $a_{2}$ to $a_{1}$. 

\section{Perfect nonreciprocity}
Besides the blockade of energy transmission in one direction, perfect nonreciprocity also requires that the input field from the opposite direction can be transmitted to the output without any insertion loss. This indicates that achieving perfect nonreciprocity corresponds to establishing an effective unidirectional lossless channel for energy transmission to guarantee $T_{\rightarrow}=1$, $T_{\leftarrow}=0$  (or $T_{\rightarrow}=0$, $T_{\leftarrow}=1$). 
To achieve lossless forward transmission ($T_{\rightarrow}=1$), as illustrated in Fig.~\ref{fig1}(c),  we engineer the interference of the two possible channels for energy transmission from the input mode to the intermediate mode ($a_{1}\rightarrow b$ and $a_{1}\rightarrow a_{2}\rightarrow b$) to be destructive. This leads to zero energy occupation on $b$ for forward transmission, which means that the  intermediate mode $b$ is effectively transparent for forward transmission. From Eq.~\eqref{eq:ssol}, this corresponds to tuning $B_{1}\propto g_{a}g_{r}e^{i\theta}+g_{l}|\Omega_{a}^{(2)}|e^{i\phi_{a}^{(2)}}$ to be 0. This can be satisfied by matching the coupling strengths and effective resonance frequency of $a_{2}$ as $g_{a}g_{r}=g_{l}|\Omega_{a}^{(2)}|$, and the phases should be tuned as $\theta-\phi_{a}^{(2)}=(2k+1)\pi$, with $\theta\neq p\pi$ and $\phi_{a}^{(2)}\neq q\pi$, where $k,p,q$ are integers. Combining with the condition of the unidirectional forward transmission [Eq.~\eqref{eq:uni-cond}], we can finally solve $T_{\rightarrow}=1$ and  $T_{\leftarrow}=0$ simultaneously and get the condition of perfect nonreciprocity (see details in Appendix C)
\begin{equation}
\begin{aligned}
g_{l}&=\sqrt{|\Omega_{a}^{(1)}\Omega_{b}|},\\
g_{r}&=\sqrt{|\Omega_{a}^{(2)}\Omega_{b}|},\\
g_{a}&=\sqrt{|\Omega_{a}^{(1)}\Omega_{a}^{(2)}|},\\
\theta\mp\phi=(2k+&1)\pi, \quad\theta\neq p\pi, \quad\phi\neq q\pi,
\end{aligned}
\label{eq:pn-cond}
\end{equation}
where $k,p,q$ are integers. Here $- (+)$ corresponds to the condition of perfect forward (backward) nonreciprocity. The loss phases should satisfy $\phi=\phi_{a}^{(i)}=\phi_{b}$, $i\in\{1,2\}$, which can be expressed in terms of the detunings and decay rates of the resonance modes as
\begin{equation}
\frac{\gamma_{1}}{\Delta_{1}+\omega}=\frac{\gamma_{2}}{\Delta_{2}+\omega}=\frac{\kappa}{\delta+\omega}.
\end{equation}


\begin{figure}
\includegraphics[width=\columnwidth]{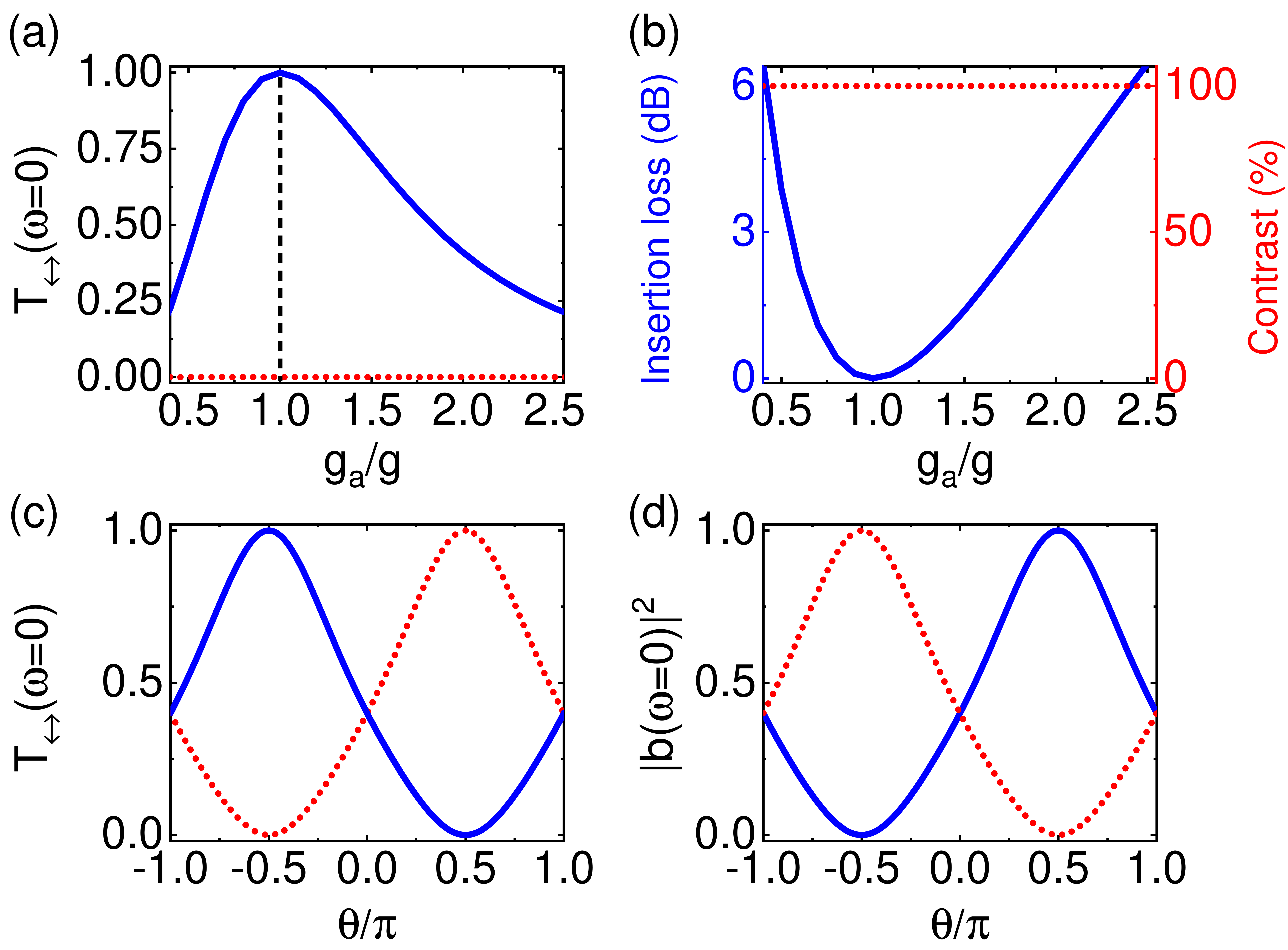}
\caption{(a) Forward  transmission efficiency ($T_{\rightarrow}$, blue solid curve) at the resonant point ($\omega=0$) as a function of the coupling strength ratio $g_{a}/g$, where $g_{l}=g_{r}=g$ is assumed for simplicity. The backward transmission efficiency under the condition of unidirectional forward transmission [Eq.~\eqref{eq:uni-cond}] is 0 ($T_{\leftarrow}$, red dotted curve). (b) The corresponding insertion loss for unidirectional forward transmission and nonreciprocity contrast as a function of the coupling strength ratio $g_{a}/g$.    (c) The  transmission efficiencies ($T_{\rightarrow}$, blue solid curve; $T_{\leftarrow}$, red dotted curve) at the resonant point ($\omega=0$) as functions of the coupling phase $\theta$, where the coupling strengths are chosen to fulfill the condition given in Eq~\eqref{eq:pn-cond}. (d) The corresponding energy occupation of the intermediate mode $b$ at the resonant point ($\omega=0$) as a function of the coupling phase $\theta$ when considering input from $a_{1}$ (blue solid curve) and $a_{2}$ (red dotted curve).
 The left parameters are fixed as  $\gamma_{1/2}=\kappa$, $\Delta_{1/2}=\delta=0$.}
\label{fig2}
\end{figure}

Figure~\ref{fig2}(a) presents the transmission efficiencies $T_{\leftrightarrow}$ as functions of the ratio of the coupling strengths ($g_{a}/g$), where we have assumed $g=g_{l}=g_{r}$ for simplicity. Under the condition of unidirectional forward transmission [Eq.~\eqref{eq:uni-cond}],  we obtain $T_{\leftarrow}(\omega)=0$. This indicates that the nonreciprocity contrast evaluated as the difference between forward and backward transmission efficiencies reaches 100\% for arbitrary coupling strength ratio $g_{a}/g$ as shown in Fig.~\ref{fig2}(b). By tuning the coupling strengths to fulfill the perfect forward nonreciprocity condition [Eq.~\eqref{eq:pn-cond}], i.e., $g_{a}/g=1$, the forward transmission efficiency reaches its maximum $T_{\rightarrow}=1$, indicating that the system exhibits zero  insertion loss for unidirectional forward transmission [Fig.~\ref{fig2}(b)]. 

The interference mechanism behind perfect nonreciprocity realization is demonstrated in Figs.~\ref{fig2}(c) and 2(d). When the coupling strengths are tuned to satisfy the condition of perfect nonreciprocity [Eq.~\eqref{eq:pn-cond}], the transmission efficiencies are solely determined by the direct coupling phase $\theta$ for fixed loss phase $\phi=\pi/2$. When tuning the phases to satisfy  $\theta=\phi-\pi=-\pi/2$, $T_{\rightarrow}=1$ (blue solid curve) and $T_{\leftarrow}=0$ (red dotted curve) can be obtained simultaneously as depicted in Fig.~\ref{fig2}(c).
This is because at this point, in addition to the destructive interference that has been tuned for backward transmission from $a_{2}$ to $a_{1}$, the direct and indirect channel for energy transmission from $a_{1}$ to the intermediate mode $b$ ($a_{1}\rightarrow b$ and $a_{1}\rightarrow a_{2}\rightarrow b$), as illustrated in Fig.~\ref{fig1}(c), also interfere destructively.  This guarantees zero energy occupation on $b$ [Fig.~\ref{fig2}(d), blue solid curve] for input from $a_{1}$, which indicates that the unidirectional channel for forward energy transmission can also be lossless. When considering input from $a_{2}$, the dual-channel destructive interference requires the phases to be matched as $\theta=\pi/2$. In this case,  zero energy occupation on $b$ makes it effectively transparent for backward transmission [Fig.~\ref{fig2}(d), red dotted curve], leading to the generation of perfect backward nonreciprocity, i.e., $T_{\rightarrow}=0$, $T_{\leftarrow}=1$, as shown in Fig.~\ref{fig2}(c).

\begin{figure}[tbp]
\includegraphics[width=\columnwidth]{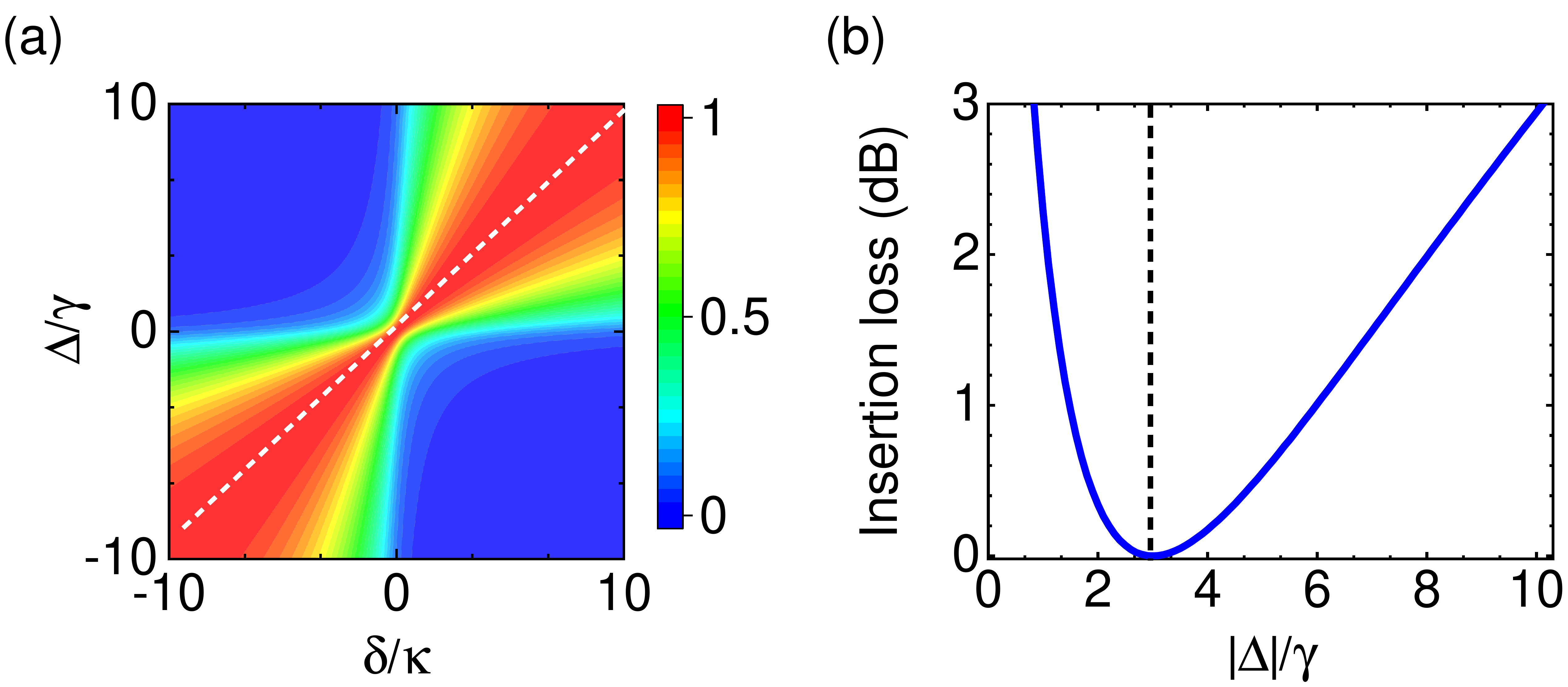}
\caption{(a) The unidirectional forward transmission efficiency $T_{\rightarrow}$ at the resonant point ($\omega=0$) as a function of the detuning-to-loss ratios $\delta/\kappa$ and $\Delta/\gamma$, where the coupling strengths are tuned to satisfy the condition of perfect nonreciprocity [Eq.~\eqref{eq:pn-cond}]. The white dashed line illustrates the condition of perfect nonreciprocity for the detuning-to-loss ratios, i.e.,  $\Delta/\gamma=\delta/\kappa$. (b) The corresponding insertion loss for unidirectional forward transmission as a function of the ratio $\Delta/\gamma$ when $\delta/\kappa=3$. }
\label{fig3}
\end{figure}

By expressing the loss phases in terms of the detunings and losses of the resonance modes,  we plot $T_{\rightarrow}(\omega=0)$ as a function of the detuning-to-loss ratios in Fig.~\ref{fig3}(a) in the case of unidirectional forward transmission,  where $\Delta=\Delta_{i}$ and $\gamma=\gamma_{i}$, $i\in\{1,2\}$, have been assumed for simplicity. When tuning the coupling strengths to satisfy the  condition of perfect forward nonreciprocity given in Eq.~\eqref{eq:pn-cond}, $T_{\rightarrow}(\omega=0)>50\%$ requires that the detunings of the modes ($a_{1}$, $a_{2}$, and $b$) have the same sign. Lossless unidirectional forward transmission $T_{\rightarrow}(\omega=0)=1$ corresponds to choosing the detuning-to-loss ratios to be equal, i.e., $\delta/\kappa=\Delta/\gamma$ [Fig.~\ref{fig3}(a), white dashed line]. To demonstrate the broad detuning range for achieving low insertion loss for unidirectional forward transmission, we fix $\delta/\kappa=3$ and the transmission insertion loss becomes a function of the ratio $\Delta/\gamma$. The insertion loss is kept below 3dB by tuning $0.9<(\Delta/\gamma)<10$ as shown in Fig.~\ref{fig3}(b).

\section{Loss effect on nonreciprocity}

\begin{figure}
\includegraphics[width=\columnwidth]{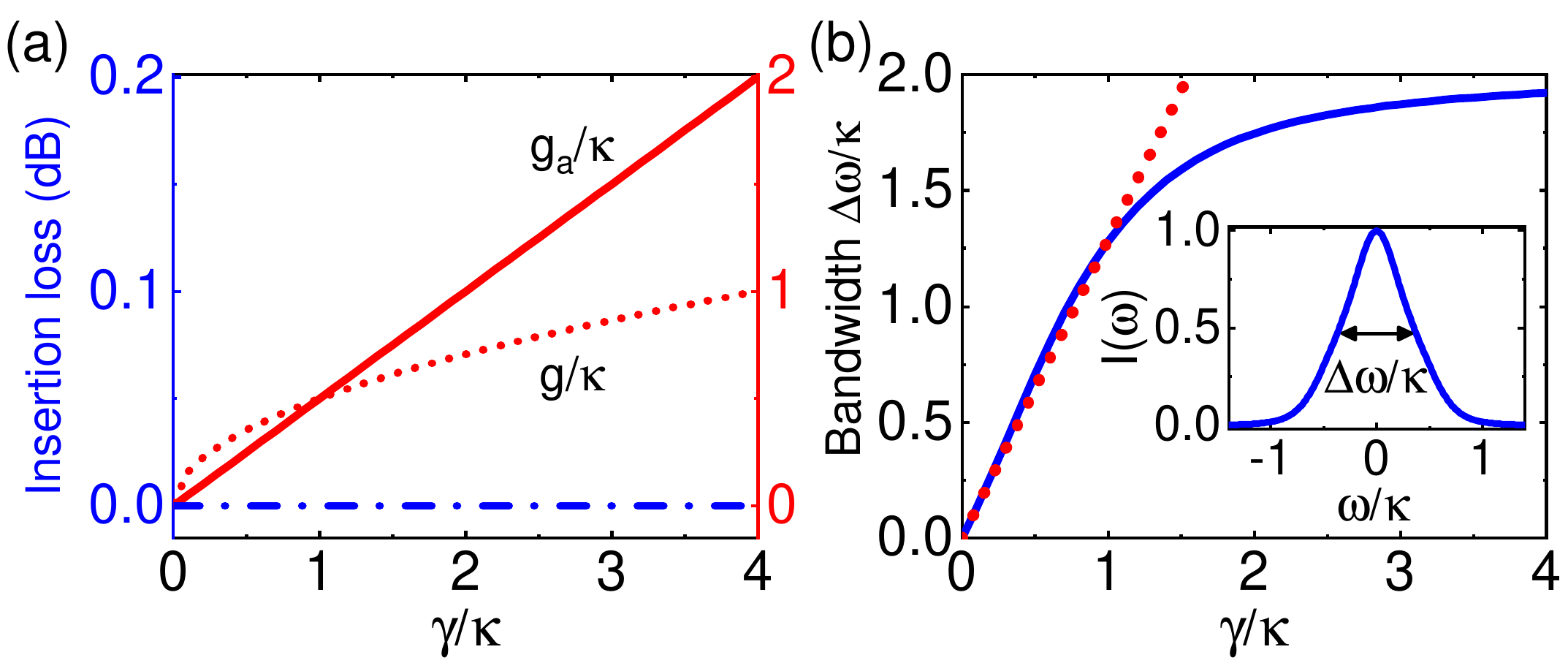}
\caption{(a) The insertion loss of the unidirectional forward transmission  and the required coupling strengths [determined by the condition of perfect forward nonreciprocity given in Eq.~\eqref{eq:pn-cond}] as functions of the energy loss rate $\gamma$. (b) Nonreciprocity bandwidth ($\Delta \omega$) as a function of the loss rate $\gamma$. The red dotted curve illustrates the asymptotic scaling of the nonreciprocity bandwidth in the limit of $\gamma/\kappa\ll1$. The inset plots the nonreciprocity spectrum function $I(\omega)=T_{\rightarrow}(\omega)-T_{\leftarrow}(\omega)$ when choosing $\gamma/\kappa=20$.}
\label{fig4}
\end{figure}

A key advantage of our scheme that we will discuss in this section is the robustness against energy loss. It can be understood from the condition of perfect nonreciprocity given in Eq.~\eqref{eq:pn-cond}, as one can enhance the coupling strengths accordingly  to keep the unidirectional transmission lossless when increasing the energy losses of the resonant modes.
To show the results more intuitively, we assume $\gamma=\gamma_{i}$, $i\in\{1,2\}$, for simplicity and plot the insertion loss of the unidirectional forward transmission as a function of the loss rate $\gamma$ in Fig.~\ref{fig4}(a).  By optimizing the coupling strengths ($g, g_{a}$), the insertion loss for unidirectional forward transmission is kept at 0 when increasing  $\gamma$. This demonstrates that the perfect nonreciprocity generated in our scheme is robust against energy loss. In addition, we also calculate the nonreciprocity bandwidth evaluated as $\Delta\omega=\omega_{+}-\omega_{-}$ by defining the spectrum function as $I(\omega)=T_{\rightarrow}(\omega)-T_{\leftarrow}(\omega)=\gamma_{1}\gamma_{2}(|C_{1}|^{2}-|A_{2}|^{2})$. When assuming $\gamma=\gamma_{i}$, $i\in\{1,2\}$, for simplicity, the nonreciprocity bandwidth is broadened with increasing loss rate $\gamma$ [Fig.~\ref{fig4}(b)], and an upper bound $\Delta \omega/\kappa\approx2$ can be obtained. The asymptotic scaling function of the nonreciprocity bandwidth can be derived as $2\sqrt{\sqrt{2}-1}\gamma$ in the limit of $\gamma\ll\kappa$ (a detailed derivation can be found in Appendix D).

\section{Implementation and Conclusion}\label{sec: conclusion}

Having completed the analysis of our generic scheme, we now propose some candidate platforms for experimental implementation. One example is based on an optomechanical setup where  two coupled optical modes via backscattering are made to interact with one mechanical mode~\cite{Chen2021}. In this system, the photon-phonon coupling strengths and phases can be tuned independently by controlling the amplitudes and phases of the external driving fields. Another platform is a dissipative Aharonov-Bohm ring on a momentum lattice, which is implemented by using multiphoton processes to couple the discrete momentum states of ultracold atoms~\cite{Gou2020}. In this system, the coupling coefficients and on-site losses can be controlled independently by adjusting the corresponding lasers. In both systems nonzero synthetic magnetic flux can be tuned in the three-mode coupling loops, which corresponds to the nonzero phase of the direct coupling required  in our scheme. By tuning the system parameters satisfying the condition [Eq.~\eqref{eq:pn-cond}] given in our scheme, these systems provide feasible and tunable platforms for perfect nonreciprocity generation.

In conclusion, lossless unidirectional energy transmission, i.e., perfect nonreciprocity, can be implemented by engineering the interference between a direct and an indirect energy transmission channel established by an intermediate mode with nonzero energy loss. 
The intermediate mode provides a phase lag induced by the energy loss, which is independent of the transmission direction. By tuning the coupling strengths and phases of the two channels to fulfill the condition of perfect nonreciprocity,  the dual-channel interference can be tuned to be destructive simultaneously for backward (forward) transmission from the rightmost (leftmost) mode to the leftmost (rightmost) mode, and forward (backward) transmission from the leftmost (rightmost) mode to the intermediate mode. In this case, 100\% nonreciprocity contrast and zero insertion loss for forward (backward) transmission can be achieved at the same time.
In addition, the nonreciprocity bandwidth can be efficiently broadened by increasing the energy losses of the resonance modes while the insertion loss for unidirectional transmission is still kept at zero at resonance. Applying our scheme to the multimode chain, the nonreciprocity generated by the interference mechanism also induces the non-Hermitian skin effect, i.e., the majority of eigenstates are localized at the boundaries breaking the conventional bulk-boundary correspondence~\cite{Lee2016,Yao2018}, which warrants further study. The robustness against energy loss indicates that our scheme  provides opportunities to implement high-efficiency nonreciprocal devices, which are demanded for one-way optical communication and information processing.\\

\begin{acknowledgements}
This work was supported by the Key-Area Research and Development Program of Guangdong Province (Grant No. 2019B030330001), the National Natural Science Foundation of China (NSFC) (Grants No. 92050110, No. 12275145, No. 91736106, No. 11674390, No. 91836302, and No. 12104252), the National Key R\&D Program of China (Grant No. 2018YFA0306504), and the China Postdoctoral Science Foundation (No. BX20190179, No. 2020M670277).
\end{acknowledgements}

\begin{appendix}

\section{Derivation of the effective Hamiltonian}
\label{appendix1}
In this appendix we will show how to implement the nonzero coupling  phase $\theta$ in the effective Hamiltonian given by Eq.~\eqref{eq:effH}. One common strategy is to induce the phase by using nonlinearity and time modulation. Taking a three-wave mixing process as an example, the original Hamiltonian including this process with an auxiliary mode $c$ can be written as 
\begin{equation}
\begin{aligned}
H&=H_{0}+H_{ab}+H_{ac}\\
H_{0}&=\omega_{a}^{(1)}a_{1}^{\dagger}a_{1}+\omega_{a}^{(2)}a_{2}^{\dagger}a_{2}+\omega_{b}b^{\dagger}b+\omega_{c}c^{\dagger}c,\\
H_{ab}&=(g_{l}a_{1}+g_{r}a_{2})b^{\dagger}+\text{H.c.},\\
H_{ac}&=g_{nl}a_{1}^{\dagger}a_{2}c+if_{p}e^{-i\omega_{p}t}c^{\dagger}+\text{H.c.},
\label{eq:oH}
\end{aligned}
\end{equation}
where $\omega_{a}^{(i)}$ $(i=1,2)$, $\omega_{b}$, and $\omega_{c}$ are the resonance frequencies of modes $a_{i}$, $b$, and $c$, respectively. $H_{ac}$ corresponds to a three-wave mixing process involving modes $a_{1}$, $a_{2}$, and $c$, where the coefficient of the nonlinear coupling is $g_{nl}$.  $f_{p}$ and $\omega_{p}$ are the complex amplitude and resonance frequency of the  continuous optical field added on the mode $c$. 
We first derive the  equation of motion of mode $c$ as
\begin{equation}
\frac{dc}{dt}=-(i\omega_{c}+\kappa_{c})c+g_{nl}a_{2}^{\dagger}a_{1}+f_{p}e^{-i\omega_{p}t}.
\label{eq:eomc}
\end{equation}
In the case that the pumping field amplitude $|f_{p}|$ is strong such that $|f_{p}|\gg g_{nl}$, we can ignore the effect of the nonlinear interaction term $g_{nl}a_{1}^{\dagger}a_{2}$ on the dynamics and treat the mode $c$ classically by replacing the operator $\hat{c}$ with the constant. The solution to Eq.~\eqref{eq:eomc} is then given by
\begin{equation}
c=\frac{f_{p}}{i(\omega_{c}-\omega_{p})+\kappa_{c}/2}e^{-i\omega_{p}t}=|\bar{c}|e^{-i(\omega_{p}t+\theta)}.
\end{equation}

We will next show that $\theta=-\arg[f_{p}/(i(\omega_{c}-\omega_{p})+\kappa_{c}/2)]$ is exactly the phase of the direct coupling between modes $a_{1}$ and $a_{2}$. Plugging Eq.~\eqref{eq:eomc} into Eq.~\eqref{eq:oH}, the system Hamiltonian in the frame rotating at
$U=e^{i\omega_{p}(a_{1}^{\dagger}a_{1}-a_{2}^{\dagger}a_{2}+b^{\dagger}b)t/2}$ is given as
\begin{equation}
\begin{aligned}
H'_{0}=&-\Delta_{1}a_{1}^{\dagger}a_{1}-\Delta_{2}a_{2}^{\dagger}a_{2}-\delta b^{\dagger}b,\\
H'_{ab}=&g_{l}a_{1}b^{\dagger}+g_{r}e^{i\omega_{p}t}a_{2}b^{\dagger}+\text{H.c.},\\
H'_{ac}=&g_{a}(e^{-i\theta}a_{1}^{\dagger}a_{2}+\text{H.c.}),\\
\end{aligned}
\end{equation}
where $g_{a}=g_{nl}|\bar{c}|$. $\Delta_{1}=\omega_{p}/2-\omega_{a}^{(1)}$, $\Delta_{2}=-\omega_{p}/2-\omega_{a}^{(2)}$, and $\delta=\omega_{p}/2-\omega_{b}$ are the detunings with respect to the resonance frequencies of the corresponding modes. 
Applying the time modulation of the coupling between $a_{2}$ and $b$ by replacing $g_{r}$ with the time-dependent factor $2g_{r}\cos\omega_{p}t$, we can make the rotating-wave approximation ignore the fast-oscillating term $\propto e^{2i\omega_{p}t}$ and finally get the effective Hamiltonian [Eq.~\eqref{eq:effH}].

\section{Steady-state solution}
\label{sec2}  

Using the Fourier transform convention $\vec{v}(t)=\int d\omega e^{-i\omega t}\vec{v}(\omega)/(\sqrt{2\pi})$, the Langevin equations in the main text have the solution in the frequency domain
\begin{equation}
\begin{aligned}
a_{1}(\omega)&=\frac{-i(g_{l}b+g_{a}e^{-i\theta}a_{2})+\sqrt{\gamma_{1}^{\text{ex}}}a_{1}^{\text{in}}}{-i(\Delta_{1}+\omega)+\gamma_{1}/2},\\
b(\omega)&=\frac{-i(g_{l}a_{1}+g_{r}a_{2})}{-i(\delta+\omega)+\kappa/2},\\
a_{2}(\omega)&=\frac{-i(g_{r}b+g_{a}e^{i\theta}a_{1})+\sqrt{\gamma_{2}^{\text{ex}}}a_{2}^{\text{in}}}{-i(\Delta_{2}+\omega)+\gamma_{2}/2},
\end{aligned}
\end{equation}
where we ignore the intrinsic losses ($\gamma_{i}=\gamma_{i}^{\mathrm{ex}}+\gamma_{i}^{(0)}\approx\gamma_{i}^{\mathrm{ex}}$, $\kappa=\kappa_{\mathrm{ex}}+\kappa_{0}\approx\kappa_{\mathrm{ex}}$) and assume the input fields coupled with the modes $a_{1}$ and $a_{2}$. Expressing the steady-state solution of the field amplitude in terms of the input field operators ($a_{1}^{\text{in}}$ and $a_{2}^{\text{in}}$), we can obtain
\begin{equation}
\begin{aligned}
a_{1}(\omega)=\sqrt{\gamma_{1}}A_{1}a_{1}^{\text{in}}+\sqrt{\gamma_{2}}A_{2}a_{2}^{\text{in}},\\
b(\omega)=\sqrt{\gamma_{1}}B_{1}a_{1}^{\text{in}}+\sqrt{\gamma_{2}}B_{2}a_{2}^{\text{in}},\\
a_{2}(\omega)=\sqrt{\gamma_{1}}C_{1}a_{1}^{\text{in}}+\sqrt{\gamma_{2}}C_{2}a_{2}^{\text{in}},
\end{aligned}
\end{equation}
with the coefficients 
\begin{equation}
\begin{aligned}
A_{1}&=\frac{i}{F}(g_{r}^{2}-\Omega_{a}^{(2)}\Omega_{b}),\\
A_{2}&=\frac{-i}{F}(g_{l}g_{r}+g_{a}e^{-i\theta}\Omega_{b}),\\
B_{1}&=\frac{-i}{F}(g_{a}g_{r}e^{i\theta}+g_{l}\Omega_{a}^{(2)}),\\
B_{2}&=\frac{-i}{F}(g_{l}g_{a}e^{-i\theta}+g_{r}\Omega_{a}^{(1)}),\\
C_{1}&=\frac{-i}{F}(g_{l}g_{r}+g_{a}e^{i\theta}\Omega_{b}),\\
C_{2}&=\frac{i}{F}(g_{l}^{2}-\Omega_{a}^{(1)}\Omega_{b}).
\end{aligned}
\end{equation}
Here, the parameters are defined as $F=g_{l}g_{r}g_{a}e^{-i\theta}+g_{l}g_{r}g_{a}e^{i\theta}+g_{l}^{2}\Omega_{a}^{(2)}+g_{r}^{2}\Omega_{a}^{(1)}+g_{a}^{2}\Omega_{b}-\Omega_{a}^{(1)}\Omega_{a}^{(2)}\Omega_{b}$, $\Omega_{a}^{(n)}=\Delta_{n}+\omega+i\gamma_{n}/2$, and $\Omega_{b}=\delta+\omega+i\kappa/2$.
 We next plug the steady-state solution of the field amplitudes into the input-output relation $\vec{v}_\mathrm{out}(\omega)=\vec{v}_\mathrm{in}(\omega)-\sqrt{\Gamma_\mathrm{ex}}\vec{v}(\omega)$. Here, the input and output operators can be simplified as $\vec{v}_\mathrm{in}=(a_{1}^\mathrm{in},  a_{2}^\mathrm{in})^{\text{T}}$ and $\vec{v}_\mathrm{out}=(a_{1}^\mathrm{out},  a_{2}^\mathrm{out})^{\text{T}}$, as the input and output fields only couple with the modes $a_{1}$ and $a_{2}$. Then we can obtain
\begin{equation}
\vec{v}_\mathrm{out}(\omega)=S(\omega)\vec{v}_\mathrm{in}(\omega).
\label{eq:I/O}
\end{equation}
The scattering matrix $S(\omega)$ can be written as
\begin{equation}
 S(\omega)=I-\begin{pmatrix} A_{1}\gamma_{1}&&A_{2}\sqrt{\gamma_{1}\gamma_{2}}\\C_{1}\sqrt{\gamma_{1}\gamma_{2}}&&C_{2}\gamma_{2} \end{pmatrix},
 \end{equation}
where $I$ is the identity matrix. Hence the forward and backward energy transmission efficiencies are given by 
\begin{equation}
\begin{aligned}
T_{\rightarrow}=|\langle a_{2}^{\text{out}}/a_{1}^{\text{in}}\rangle|^{2}=|S_{21}(\omega)|^{2}=\gamma_{1}\gamma_{2}|C_{1}|^{2},\\
T_{\leftarrow}=|\langle a_{1}^{\text{out}}/a_{2}^{\text{in}}\rangle|^{2}=|S_{12}(\omega)|^{2}=\gamma_{1}\gamma_{2}|A_{2}|^{2},
\label{eq:T}
\end{aligned}
\end{equation}
respectively.

\section{Condition of perfect nonreciprocity}
\label{sec3}

To achieve perfect nonreciprocity, i.e., lossless unidirectional energy transmission, the interference between the direct and indirect channels for unidirectional energy transmission should be completely constructive, which means that $|C_{1}|=1/\sqrt{\gamma_{1}\gamma_{2}}$ ($|A_{2}|=1/\sqrt{\gamma_{1}\gamma_{2}}$) for unidirectional forward (backward) transmission. Noting that the complete energy transfer  from $a_{1}$ ($a_{2}$) to $a_{2}$ ($a_{1}$) corresponds to the zero energy occupation of the intermediate mode $b_{1}$, we can then obtain the hidden condition of perfect nonreciprocity, i.e., the destructive interference $C_{1}=0$ ($C_{2}=0$) between the two possible channels for energy transmission from the input mode $a_{1}$ ($a_{2}$) to the intermediate mode $b$. Let us first consider the forward transmission. The relation of the coupling strengths is found to be 
 \begin{equation}
 g_{a}g_{r}/g_{l}= |\Omega_{a}^{(2)}|,
 \end{equation} 
and the phases need to match as $\theta-\phi_{a}^{(2)}=(2k+1)\pi$, with $\theta\neq p\pi$ and $\phi_{a}^{(2)}\neq q\pi$, where $k,p$, and $q$ are integers. Combining this with the condition of unidirectional forward transmission, we can have
\begin{equation}
g_{r}=\sqrt{|\Omega_{a}^{(2)}\Omega_{b}|}, \phi_{a}^{(2)}=\phi_{b}.
\label{eq:dc}
\end{equation}

Plugging Eq.~\eqref{eq:dc} into the complete constructive interference condition for forward transmission ($|B_{1}|=1/\sqrt{\gamma_{1}\gamma_{2}}$), we can get
\begin{widetext}
\begin{equation}
(g_{a}^{2}-(\Delta_{1}+\omega)(\Delta_{2}+\omega)-\frac{\gamma_{1}\gamma_{2}}{4})^{2}+\frac{1}{4}(\gamma_{2}(\Delta_{1}+\omega)-\gamma_{1}(\Delta_{2}+\omega))^{2}=0,
\end{equation}
\end{widetext}
which can be fulfilled only when 
\begin{equation}
\begin{aligned}
\frac{\gamma_{1}}{\Delta_{1}+\omega}=\frac{\gamma_{2}}{\Delta_{2}+\omega},\\
g_{a}=\sqrt{|\Omega_{a}^{(1)}\Omega_{a}^{(2)}|}.
\end{aligned}
\end{equation}

Hence perfect forward nonreciprocity can be achieved when tuning the coupling strengths to be
\begin{equation}
\begin{aligned}
g_{l}&=\sqrt{|\Omega_{a}^{(1)}\Omega_{b}|},\\
g_{r}&=\sqrt{|\Omega_{a}^{(2)}\Omega_{b}|},\\
g_{a}&=\sqrt{|\Omega_{a}^{(1)}\Omega_{a}^{(2)}|}
\end{aligned}
\end{equation}
and matching the coupling and loss phases as $\theta-\phi=(2k+1)\pi$, where $\phi=\phi_{a}^{(i)}=\phi_{b}$, $i\in\{1,2\}$. The condition of the loss phases can also be expressed in terms of the detunings and loss rates of the resonance modes as
\begin{equation}
\frac{\gamma_{1}}{\Delta_{1}+\omega}=\frac{\gamma_{2}}{\Delta_{2}+\omega}=\frac{\kappa}{\delta+\omega}.
\end{equation}

Similar calculations can also be performed for backward transmission. In this case, perfect backward nonreciprocity can be achieved when matching the coupling and loss phases as $\theta+\phi=(2k+1)\pi$.

\section{Nonreciprocity bandwidth}
\label{sec4}

To calculate the bandwidth of nonreciprocity, we first define the spectrum function as $I(\omega)=T_{\rightarrow}(\omega)-T_{\leftarrow}(\omega)=\gamma_{1}\gamma_{2}(|C_{1}|^{2}-|A_{2}|^{2})$.  Considering the case of  perfect forward nonreciprocity at resonance, i.e., $T_{\rightarrow}(\omega=0)=1$, $T_{\leftarrow}(\omega=0)=0$, and choosing $\delta=\Delta_{i}=0$, $i\in\{1,2\}$, for simplicity,  the spectrum function can be derived as
\begin{widetext}
\begin{equation}
I(\omega)=\frac{(\gamma_{1}^{2}\gamma_{2}^{2}/4)(|\omega+i\kappa|^{2}-|\omega|^{2})}{|(\gamma_{1}\kappa/4)(\omega+i\gamma_{2}/2)+(\gamma_{2}\kappa/4)(\omega+i\gamma_{1}/2)+(\gamma_{1}\gamma_{2}/4)(\omega+i\kappa/2)-(\omega+i\gamma_{1}/2)(\omega+i\gamma_{2}/2)(\omega+i\kappa/2)|^{2}}.
\label{eq:sf}
\end{equation}
\end{widetext}

In the general case of $\gamma_{1}=a\gamma_{2}$, Eq.~\eqref{eq:sf} can be further simplified as
\begin{widetext}
\begin{equation}
I(\omega')=\frac{a^{2}\gamma'^{4}}{4\omega'^{6}+\omega'^{4}[(a-1)^{2}\gamma'^{2}-2(a+1)\gamma'+1]+\omega'^{2}\gamma'^{2}(a^{2}\gamma'^{2}+a^{2}+1)+a^{2}\gamma'^{4}},
\end{equation}
\end{widetext}
where $\omega'=\omega/\kappa$ and $\gamma'=\gamma_{2}/\kappa$. Hence the solution to $I(\omega'_{\pm})=1/2$ corresponds to the nonreciprocity bandwidth $\Delta\omega/\kappa=\omega'_{+}-\omega'_{-}$.

For quantitative understanding of the relation between the bandwidth and the loss rates, we now derive the asymptotic scaling function of the bandwidth  in the limit of $\gamma'\ll1$. It can be calculated by solving the bandwidth equation
\begin{widetext}
\begin{equation}
4\omega'^{6}+\omega'^{4}[(a-1)^{2}\gamma'^{2}-2(a+1)\gamma'+1]+\omega'^{2}\gamma'^{2}(a^{2}\gamma'^{2}+a^{2}+1)-a^{2}\gamma'^{4}=0
\label{eq:bandwidth}
\end{equation}
\end{widetext}
and performing Taylor expansion for the solution $\Delta\omega/\kappa=\omega'_{+}-\omega'_{-}$ in the limit of $\gamma'\ll1$. The leading contribution is derived as
\begin{equation}
\Delta\omega\approx \sqrt{2}\cdot\sqrt{-(a^{2}+1)+\sqrt{a^{4}+6a^{2}+1}}\gamma_{2}.
\end{equation}

When assuming $\gamma_{1}=\gamma_{2}=\gamma$, i.e., $a=1$, the asymptotic scaling function can be simplified as
\begin{equation}
\Delta\omega\approx2\sqrt{\sqrt{2}-1}\gamma.
\end{equation}

This indicates that the nonreciprocity bandwidth is broadened linearly with increasing loss rate $\gamma$ in the limit of $\gamma\ll\kappa$.  Maximizing the solution to Eq.~\eqref{eq:bandwidth} and performing Taylor expansion in the limit of $\gamma'\gg1$, we find the maximum of the nonreciprocity bandwidth $\Delta \omega/\kappa\approx2$, which is independent of $\gamma_{1}$ and $\gamma_{2}$.
\end{appendix}

\bibliographystyle{apsrev4-1}
\bibliography{ref-PerfectNonreciprocity.bib}

\end{document}